\documentclass{wscpaperproc}
\usepackage{latexsym}
\usepackage{caption}
\usepackage{subcaption}
\usepackage{graphicx}
\usepackage{mathptmx}

\usepackage{amsmath}
\usepackage{amsfonts}
\usepackage{amssymb}
\usepackage{amsbsy}
\usepackage{amsthm}

\usepackage[pdftex,colorlinks=true,urlcolor=blue,citecolor=black,anchorcolor=black,linkcolor=black]{hyperref}

\newtheoremstyle{wsc}
{3pt}
{3pt}
{}
{}
{\bf}
{}
{.5em}
{}

\theoremstyle{wsc}

\begin{document}

%
%

\pagestyle{fancyplain}

\thispagestyle{plain}
\firstPageHead{}

\chead{\fancyplain{}{\itshape Diamantopoulos, Tziritas, Bahsoon and Theodoropoulos}}

\rhead{}
\cfoot{}
\renewcommand{\headrulewidth}{0pt} 

\makeatletter
\let\@internalcite\cite
\def\cite{\def\@citeseppen{-1000}%
    \def\@cite##1##2{(##1\if@tempswa , ##2\fi)}%
    \def\citeauthoryear##1##2##3{##1 ##3}\@internalcite}
\def\citeNP{\def\@citeseppen{-1000}%
    \def\@cite##1##2{##1\if@tempswa , ##2\fi}%
    \def\citeauthoryear##1##2##3{##1 ##3}\@internalcite}
\def\citeN{\def\@citeseppen{-1000}%
    \def\@cite##1##2{##1\if@tempswa, ##2)\else{}\fi}%
    \def\citeauthoryear##1##2##3{##1 (##3)}\@citedata}
\def\citeA{\def\@citeseppen{-1000}%
    \def\@cite##1##2{(##1\if@tempswa , ##2\fi)}%
    \def\citeauthoryear##1##2##3{##1}\@internalcite}
\def\citeANP{\def\@citeseppen{-1000}%
    \def\@cite##1##2{##1\if@tempswa , ##2\fi}%
    \def\citeauthoryear##1##2##3{##1}\@internalcite}
\def\shortcite{\def\@citeseppen{-1000}%
    \def\@cite##1##2{(##1\if@tempswa , ##2\fi)}%
    \def\citeauthoryear##1##2##3{##2 ##3}\@internalcite}
\def\shortciteNP{\def\@citeseppen{-1000}%
    \def\@cite##1##2{##1\if@tempswa , ##2\fi}%
    \def\citeauthoryear##1##2##3{##2 ##3}\@internalcite}
\def\shortciteN{\def\@citeseppen{-1000}%
    \def\@cite##1##2{##1\if@tempswa, ##2\else{}\fi}%
    \def\citeauthoryear##1##2##3{##2 (##3)}\@citedata}
\def\shortciteA{\def\@citeseppen{-1000}%
    \def\@cite##1##2{(##1\if@tempswa , ##2\fi)}%
    \def\citeauthoryear##1##2##3{##2}\@internalcite}
\def\shortciteANP{\def\@citeseppen{-1000}%
    \def\@cite##1##2{##1\if@tempswa , ##2\fi}%
    \def\citeauthoryear##1##2##3{##2}\@internalcite}
\def\citeyear{\def\@citeseppen{-1000}%
    \def\@cite##1##2{(##1\if@tempswa , ##2\fi)}%
    \def\citeauthoryear##1##2##3{##3}\@citedata}
\def\citeyearNP{\def\@citeseppen{-1000}%
    \def\@cite##1##2{##1\if@tempswa , ##2\fi}%
    \def\citeauthoryear##1##2##3{##3}\@citedata}
%
%
%
\def\@citedata{%
    \@ifnextchar [{\@tempswatrue\@citedatax}%
                  {\@tempswafalse\@citedatax[]}%
}

\def\@citedatax[#1]#2{%
\if@filesw\immediate\write\@auxout{\string\citation{#2}}\fi%
  \def\@citea{}\@cite{\@for\@citeb:=#2\do%
    {\@citea\def\@citea{, }\@ifundefined
       {b@\@citeb}{{\bf ?}%
       \@warning{Citation `\@citeb' on page \thepage \space undefined}}%
{\csname b@\@citeb\endcsname}}}{#1}}%

%
\def\@citex[#1]#2{%
\if@filesw\immediate\write\@auxout{\string\citation{#2}}\fi%
  \def\@citea{}\@cite{\@for\@citeb:=#2\do%
    {\@citea\def\@citea{; }\@ifundefined
       {b@\@citeb}{{\bf ?}%
       \@warning{Citation `\@citeb' on page \thepage \space undefined}}%
{\csname b@\@citeb\endcsname}}}{#1}}%

%
\def\@biblabel#1{}
\makeatother



\newdimen\bibindent
\bibindent=0.0em
\def\thebibliography#1{\section*{\refname}\list
   {}{\settowidth\labelwidth{[#1]}
   \leftmargin\parindent
   \itemindent -\parindent
   \listparindent \itemindent
   \itemsep 0pt
   \parsep 0pt}
   \def\newblock{}
   \sloppy
   \sfcode`\.=1000\relax}


\setlength{\baselineskip}{12.7pt}

\title{DIGITAL TWINS FOR DYNAMIC MANAGEMENT OF BLOCKCHAIN SYSTEMS}

\author{Georgios Diamantopoulos\\[12pt]
    Department of Computer Science and Engineering\\
Southern University of Science and Technology\\
Shenzhen, China \\
and\\
School of Computer Science\\
    University of Birmingham\\
    Birmingham, United Kingdom\\
\and
Nikos Tziritas\\[12pt]
Department of Informatics\\ and Telecommunications\\
University of Thessaly\\
Greece\\
\and
Rami Bahsoon\\ [12pt]
School of Computer Science\\
University of Birmingham\\
United Kingdom\\
\and
Georgios Theodoropoulos\\[12pt]
Department of Computer Science and Engineering\\
Southern University of Science and Technology\\
Shenzhen, China\\
}

\maketitle

\section*{ABSTRACT}
Blockchain systems are challenged by the so-called Trilemma tradeoff: decentralization, scalability and
security. Infrastructure and node configuration, choice of the Consensus Protocol and complexity of the
application transactions are cited amongst the factors that affect the tradeoffs balance. Given that Blockchains
are complex, dynamic dynamic systems, a dynamic approach to their management and reconfiguration at
runtime is deemed necessary to reflect the changes in the state of the infrastructure and application. This
paper introduces the utilisation of Digital Twins for this purpose. The novel contribution of the paper is design
of a framework and conceptual architecture of a Digital Twin that can assist in maintaining the Trilemma
tradeoffs of time critical systems. The proposed Digital Twin is illustrated via an innovative approach
to dynamic selection of Consensus Protocols. Simulations results show that the proposed framework can
effectively support the dynamic adaptation and management of the Blockchain.

\section{INTRODUCTION}

Blockchain has seen a huge leap in popularity since its inception as an immutable, decentralised ledger used by Bitcoin~\cite{btc} and the plethora of other applications that soon followed. In Blockchain entities that wish to transact with each other form a P2P network through which cryptographically signed transactions are batched into blocks, broadcasted, and stored in a chain of blocks by every entity individually. A transaction is defined as the transfer of a digital token which can be designed to virtually any functionality through a process called tokenization~\cite{tokenization}. A major distinction between Blockchains is their type, with the two main categories being permissionless or public and permissioned or private~\cite{publicVprivate}; a consortium Blockchain~\cite{consortium} is a hybrid type encopassing features of both main ones. In a permissionless Blockchain the P2P network is public and everyone can participate anonymously. As a result, the network topology is unknown and no a-priori assumptions can be made about nodes or expected load of the system. In the permissioned case, the P2P network is private and only verified nodes can participate thus providing more knowledge about the state of the system. Finally in a consortium Blockchain, the network is public and every one can participate but only verified nodes can produce blocks. 

Blockchain has been increasingly utilised in a wide range of applications including IoT, supply chain systems, e-government systems, medical databases
and more recently metaverse type applications~\cite{doi:10.1504/IJWGS.2018.095647,8662573,DIFRANCESCOMAESA202099,8731639,8805074,Gamage,metaverse}.
The potential of Blockchain technology to support sustainable development is also increasingly being acknowledged while tokenisation is viewed as the key technology to promote and power ESG, impact investment and sustainable finance~\cite{sustainability,UNCTAD}. 


Despite the widely acknowledged potentials of Blockchain, there are several factors that limit, if not prohibit its adoption in time critical applications: low scalability, high latency, coupled with high power consumption, and an expanding carbon footprint are among the most cited factors~\cite{scalabIssue}. As an indicative example, Bitcoin can confirm an average of 4 transactions per second (TPS) and Ethereum's public implementation can confirm an average of 14 TPS\footnote{\url{https://www.Blockchain.com/charts}}; 
in comparison, VISA, a traditional transaction processing system, claims to process more than 24,000 TPS\footnote{\url{https://usa.visa.com/run-your-business/small-business-tools/retail.html}}. 
Henceforth, the designers of Blockchain-based systems are pressured by the need to develop secure, scalable, speedy and sustainable solutions.

Aspiring to contribute to this endeavour, this paper presents an approach for the dynamic management and optimisation of permissioned Blockchain systems utilising Digital Twins. The novel contribution of this paper is the design
of a framework and  a conceptual architecture leveraging Digital Twin technology to assist application designers in maintaining the so called Trilemma
tradeoff in Blockchain-based systems (coined
as suggested by Ethereum's Vitalik Buterin): decentralisation, scalability and security. 


Our approach views Digital Twin as a ``combination of a computational model and a real-world system, designed to monitor, control and optimise its functionality"\cite{arup}. The objective of using Digital Twins is to dynamically assist in managing and optimising for the Trilemma tradeoffs in Blockchain-based systems. Our approach is fundamentally grounded on the premise that Digital Twins are essentially Dynamic Data Driven Application Systems (DDDAS), wherein a real-time info-symbiotic feedback loop between the model and the real system allows data from an observed system to be absorbed into a simulation of the system in order to continually adapt the model to the reality, if necessary making changes to the assumptions on which it is based in order to gradually increase the reliability of its forecasts. Additionally, the predictions of the simulation can be fed back to the observed system in order to change or optimise its behaviour in real time and direct the data collection and sampling~\cite{10.1007/978-3-540-69389-5_2}.

Digital Twins and DDDAS have been utilised in a wide range of applications~\cite{dddas,Jones2020CharacterisingReview,Minerva2020,DosSantos2021,Barricelli2019}, including autonomic management of computational infrastructures~\cite{10.5555/2429759.2429956,ONOLAJA20101241,FANIYI20121167,6838768}.    
The last few years have witnessed several efforts to bring together Blockchain and Digital Twins, however these have focused on utilising the former to support the latter; a comprehensive survey is provided in ~\cite{10.1145/3517189}. 
Similarly, in the context of Dynamic Data Driven Application Systems (or DDDAS),  Blockchain technology has been utilised to support different aspects of DDDAS operations and components~\cite{9004727,Ronghua,Ronghua1}.
In contrast, this paper aims to address the reverse challenge namely how can the DDDAS paradigm and Digital Twin technology be utilised to support the dynamic management and optimisation of blochckain systems.


The novel contributions of the paper are the following:
\begin{enumerate}
\item The first to propose the utilisation of Digital Twins to dynamically manage the Trilemma Tradeoffs in Blockchain systems.
\item It develops a generic reference architecture of Digital Twins for managing the Trilemma in Blockchain systems.
\item It demonstrates how the architecture can be instantiated to optimise for performance and to inform the dynamic selection and management of consensus in Blockchain-based systems.
\item Presents a quantitative analysis of dynamically  Consensus Protocols to optimise performance.
\end{enumerate}

The rest of the paper is structured as follows: Section \ref{MANAGING Blockchain DYNAMICS} discusses the factors affecting the performance of Blockchain systems and their dynamic management and the challenge of managing them. Section \ref{A REFERENCE ARCHITECTURE} presents a reference architecture of a Digital Twin for permissioned Blockchain systems outlining its main components. Section \ref{AN INSTANTIATION FOR DYNAMIC CONSENSUS MANAGEMENT} illustrates an example instantiation of the reference architecture for the dynamic management of Consensus Protocols while section \ref{EVALUATION} presents a quantitative analysis.  Finally, section \ref{CONCLUSIONS} concludes the paper and outlines paths for future research.

\section{MANAGING BLOCKCHAIN  DYNAMICS}
\label{MANAGING Blockchain DYNAMICS}

The design of Blockchain-based systems is challenged by the well known Trilemma tradeoff, coined by Vitalik Buterin, the co-founder of Ethereum: decentralization, scalability and security. Factors that affect the behaviour of the Blockchain and change the balance between these three attributes relate to computational infrastructure and node configuration, the Consensus Protocol and the complexity of the application transactions. Parameters such as network topology, bandwidth and latency, CPU and storage capacity, mining power utilization, number of nodes, distribution of mining power, block size, block interval, number of block producers, orphaning/fork probability determine the transaction throughput and energy profile of the Blockchain system ~\cite{10.1145/3297662.3365818,10.1007/978-3-030-34083-4_8,9133427,10.1007/978-3-662-58387-6_24,8436042,Klarman}. 


 Consensus Protocol is at the core of influencing the Trilemma trade off of Blockchain-based systems. In the field of distributed systems, consensus algorithms have been thoroughly studied and  optimised variants have been proposed~\cite{paxos,raft}. These variants have been generally effective in small scale systems and can be best suited for permissioned or consortium Blockchain based systems; their application to permissionless cases is not straightforward. This is attributed to the fact that permissioned/consortium Blockchain systems require a relatively smaller number of selected nodes to be in charge of producing blocks, where classical consensus algorithms can be effective. As the complexity of applications benefiting from Blockchain increases, several Consensus Protocols have been proposed with the goal of improving efficiency, scalability, transaction throughput and convergence. However, providing solutions which  maintain consistent performance over varying workloads, and in the face of changing environmental conditions and parameters remain a challenge~\cite{Consensus}. The challenge calls for dynamic and adaptive consensus to better address the Trilemma Tradeoffs in Blockchain-based systems. The concept of dynamic adaptation of consensus algorithms is further discussed in section \ref{AN INSTANTIATION FOR DYNAMIC CONSENSUS MANAGEMENT}. 

 With regard to the application transactions, smart contract systems are essentially complex systems with nonlinear profiles and emergent properties; their impact on the performance on the Blockchain system can not be determined a priori~\cite{kim,https://doi.org/10.1111/meta.12266,10.1007/978-3-030-13929-2_14,hybCharact}. The increasing complexity of transactions, partially attributed to smart contracts logic-validation, has observable impact on the performance of Blockchain-based systems.


Dynamic approach to the management and reconfiguration at runtime is deemed necessary to reflect on changes in the state of the infrastructure and application. Efforts in this direction have already commenced, looking at different aspects of  Blockchain systems such as selection of neighbor nodes~\cite{BANIATA202275} and optimisation techniques for revenue maximisation~\cite{1547-5816_2021_4_1845}.  

In ~\cite{rl} a framework in which a Reinforcement Learning (RL) agent is used to optimise a Blockchain system is proposed. The agent is tasked to solve a constraint optimisation task, that is, minimising latency while not compromising on decentralisation. This work provides a useful optimisation exercise with some interesting insights about the ability of the agent to select the best algorithm for the state provided. However, as is typical of RL, the agent is trained on historical data and cannot provide a nonlinear extrapolation of future scenarios, which is essential when modelling complex systems (as is the case of smart contract systems). 

A Digital Twin can overcome these deficiencies as its simulation infrastructure can allow what-if analysis and can act as a surrogate to explore alternative future scenarios~\cite{10.1145/2769458.2769484,10.5555/2874916.2874964}. It can also  provide support for the dynamic off-chain simulation and evaluation of smart contracts systems ~\cite{kim,info11010034,HU2021100179,Yeonsoo}. The smart contract system can then be executed off-chain in the Digital Twin environment or uploaded to the Blockchain system  thus supporting a hybrid on/off-chain execution model~\cite{https://doi.org/10.1002/cpe.5811}.

\section{TWINNING A BLOCKCHAIN}
\label{A REFERENCE ARCHITECTURE}

Blockchain is a distributed ledger technology which allows for trustless interactions between entities without a trusted middleman. Blockchain achieves the above by keeping a completely distributed and immutable ledger which stores ownership data of tokens representing physical or digital entities. A transaction in the Blockchain is defined as the change of ownership of an existing token or the generation and the assignment of ownership of a new token. In Blockchain, nodes connect with each other by forming a peer-to-peer (P2P) network and any node which wishes to send a token to another node, creates a transaction and broadcasts it over the network. Asymmetric cryptography
is used to prove the identity of nodes, by requiring every transaction and message sent in the Blockchain to be signed by a nodes private key for identification.
\begin{figure}[t]
\includegraphics[width=0.4\textwidth]{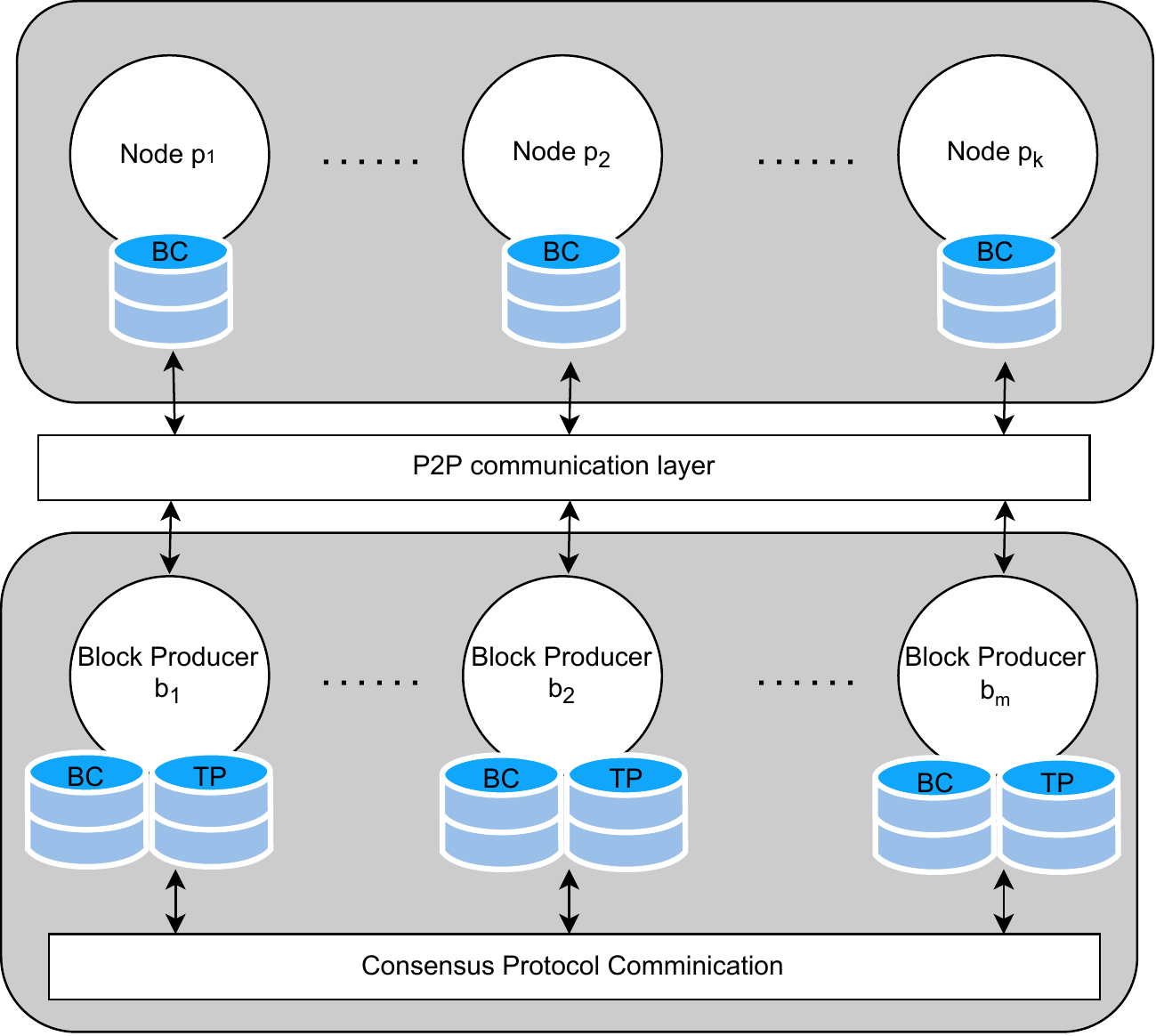}
\centering
\caption{Architecture of a generic permissioned Blockchain system} 
\label{bc_arch}
\end{figure}
When enough transactions are gathered, special nodes called block producers, batch the transactions into a block and broadcast it in the network. This block directly references the previous block creating a back-linked chain of blocks. For a new block to be valid and accepted by the rest of the nodes it needs to have been agreed upon by the Consensus Protocol. A Consensus Protocol acts as a voting mechanism in which the block producer vote on candidate blocks to be added next to the Blockchain.

In this paper, we consider a generic permissioned Blockchain system with K nodes denoted as: 

\begin{equation}
    P = \{p_1, p_2, ..., p_K\}
\end{equation}

M of which are block producers denoted as:

\begin{equation}
    B = \{b_1, b_2, ..., b_M\},  \ B \subset P
\end{equation}

\noindent which take part in the Consensus Protocol and are responsible for producing the blocks. Fig. \ref{bc_arch} illustrates the described Blockchain system. Each node $p\in P$ holds a local copy of the Blockchain (BC). Additionally, block producers  $b\in B$ also hold a transaction pool (TP) which stores broadcasted transactions that took place in the system. When a node is ready to propose a new block, the oldest transactions from the TP are selected first to populate it. Finally, when a new block is accepted the transactions included in it are removed from the node's local TP.

\subsection{A Reference Architecture}
\begin{figure}[t]
\includegraphics[width=0.9\textwidth]{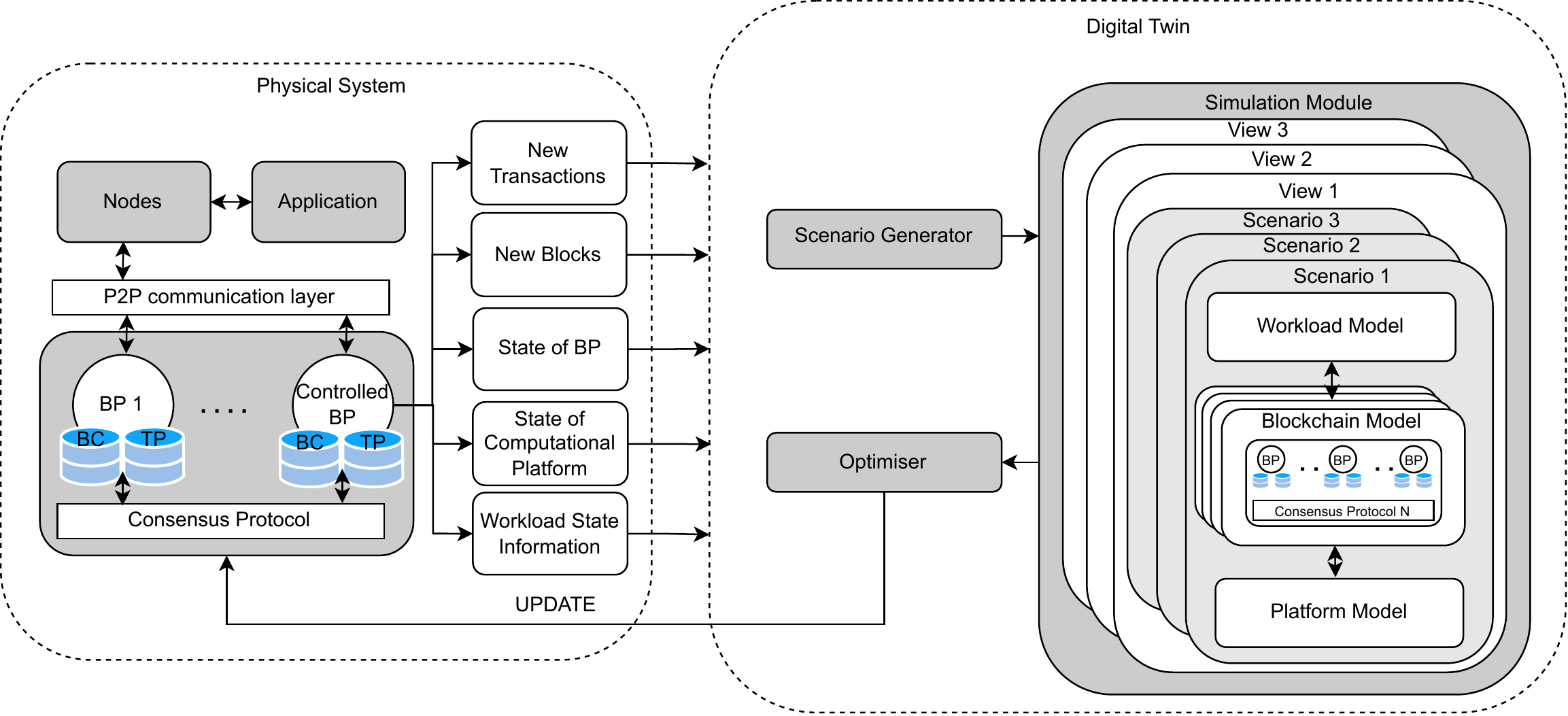}
\centering
\caption{A  reference model for a Digital Twin managed Blockchain} 
\label{dt-ref-arch}
\end{figure}

A generic reference model for the proposed Digital Twin managed Blockchain is illustrated in figure \ref{dt-ref-arch}. 
The model implements a typical MAPE-K approach~\cite{1160055}.  Following the basic philosophy of the DDDAS paradigm, data from the Blockchain system is fed to the Digital Twin at selected time intervals. The data deemed necessary to create and update the replica of the Blockchain include the following:  (a) a list of transactions received in the interval (b) a list of all new blocks added since in the interval (c) the state of the block producers (d) the state of of the computational platform and workload state information as appropriate. A new block may conatin the following: (a) the transactions included in the block (b) the list of block producers (c) the Consensus Protocol (CP) used to mine the block (d) the list of timestamped validator votes for a block (CP dependent) (e) votes to remove block producer rights from node (f) votes proposing new block producers.

Due to the decentralised nature of the Blockchain, connecting the system with it's digital representation is not a straight forward process. Unlike traditional centralised systems, with known and high speed network topologies, Blockchain's P2P network infrastructure poses a challenge in data collection. Information about nodes cannot be easily requested and aggregated. Additionally, in most Blockchain applications, nodes are assumed to be byzantine and thus any non-validated information is assumed to be malicious which further complicates data collection. One approach, proposed in this paper, is to take advantage of the verifiable transaction data broadcasted to the network and the frequent communication between the block producers as part of the consensus process, assigning a single block producer as the data provider to the Digital Twin.

The Digital Twin part encompasses three main components: The Scenario generator,  The Simulator, and the Optimiser. 
The scenario generation module, can be viewed as a high level model of the system nodes tasked with producing hypothetical workloads.


The scenarios will be fed to the Simulator which is at the heart of the Digital Twin.  This may encapsulate different data-driven models to support a holistic, contextual analysis of the system, including: (a) a model of the Blockchain system and associated infrastructure (b) agent-based models of smart contract systems (c) models of the context, e.g. in the case of a Blockchain in the energy sector, this could be models of trading, models of the regulatory and compliance framework and  a model of the energy supply chain~\cite{ANDONI2019143}. 
The simulator executes faster than real time multiple what-if scenarios for different views of the system, each view exploring an abstract aspect of the system to optimise for, for instance an energy view, a trust view, a performance view etc.


The final component is the optimiser, which is responsible for evaluating the simulation results, and selecting the best strategy under the optimisation goals. Pareto fronts and knee points analysis may be utilised to analyse the different tradeoffs involved (e.g. the cost of adaptation vs the sort and long term benefits) and make a decision as to what is the best strategy to reconfigure the Blockchain. 
The results of the simulation can be used to enhance the training an intelligent optimiser. 
In~\cite{9283357,TOMACS} we have discussed the design of intelligent Digital Twins and have presented an analysis of the tradeoffs for the adaptation of Digital Twins of agent-based systems. 


The completion of the feedback loop, namely the communication and application of the optimiser output back top the Blockchain system presents an interesting challenge. One approach is to communicate the outcome  to the entrusted controlled BP and allow this node to propagate it to the rest of the network. This may be achieved by piggybacking the information in the next block to be forwarded or through a broadcast to all other nodes.


\subsection{An Instantiation for Dynamic Consensus Management}
\label{AN INSTANTIATION FOR DYNAMIC CONSENSUS MANAGEMENT}

Given the centrality of the Consensus Protocol in the bevaviour of Blockchain systems,  as an illustrative example, this section considers the application of the reference Digital Twin model for the dynamic management of the Blockchain Consensus Protocol. As discussed in section \ref{MANAGING Blockchain DYNAMICS}, each of the existing protocols seem to work well under certain Blockchain configurations and workload conditions while none is able to deliver a consistently good general solution~\cite{Consensus,hybCharact}. 

Hybrid Consensus Protocols aim to combine elements from the two main algorithms of Proof-of-Work (PoW) and Proof-of-Stake (PoS).  In~\cite{forkFreePoA} Proof-Of-Activity is proposed which aims to improve the security of the Blockchain, however it further increases the latency and the energy consumption of the Consensus Protocol. 
In~\cite{PoWverif} a hybrid of PoS and PoW is proposed in which the block production is done through PoS but a group of validators periodically produce PoW blocks verifying the Blockchain and preventing roll backs due to forks thus increasing the performance of the Blockchain in those cases. Both the aforementioned examples designed for permissionless Blockchains.  The same applies to permissioned blockhain systems such as EOS~\cite{eos} and Hyperledger Fabric~\cite{hyper} which use traditional Byzantine Fault Tolerance (BFT) protocols~\cite{pbft,quorum,zyz}. Hybrid algorithms for permissioned systems have also been investigated~\cite{Pengxin}.

\begin{figure*}[!t]
\includegraphics[width=0.9\textwidth]{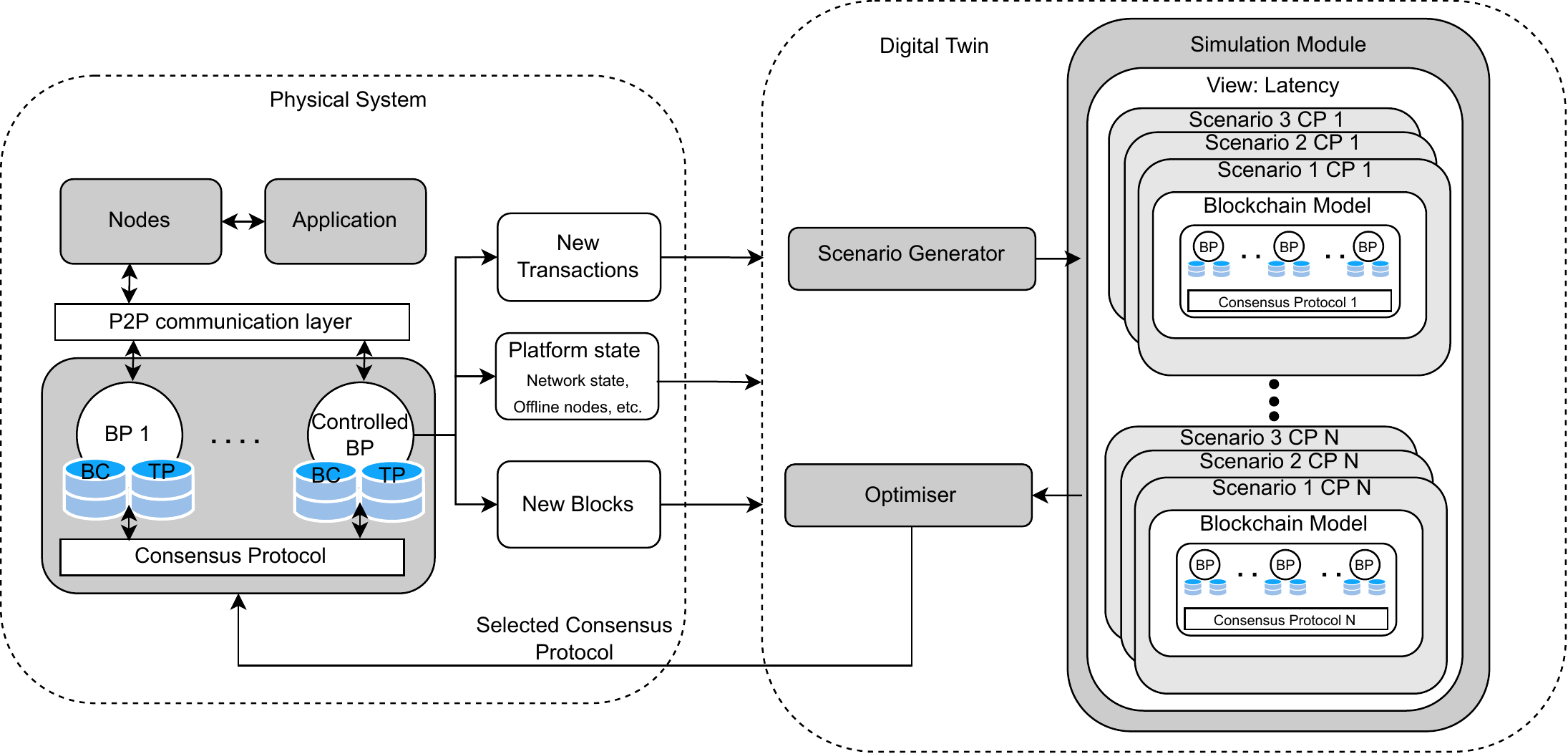}
\centering
\caption{A Digital Twin for Dynamic Consensus Protocol Selection}
\label{arch}
\end{figure*}

While hybrid algorithms try to exploit the comparative advantages of different protocols, they fail to reflect  dynamic changes of the Blockchain and the associated workloads.  it is therefore desirable for the consensus mechanism to adapt dynamically and switch to the appropriate protocol. Figure \ref{arch} illustrates the architecture of the Digital Twin for this purpose. In this particular example, the aim is the optimisation of the system's latency.

The optimisation process begins with the new transactions and new blocks being fed into the Digital Twin. The number of offline nodes and the network state may be extracted from the blocks. Specifically, offline nodes can be inferred by the lack of block votes from a particular node  while the network delay is calculated individually for every node as the average delay of their votes. 
Given the above,  the simulation conducts a what-if exploration of different scenarios for the different Consensus Protocols to predict the average transaction latency for different configurations. The transaction latency is defined as the time it takes from the moment a transaction is broadcasted to the system to the movement that transaction is packed into a block which gets accepted by the system. 
The results of the simulation module are into the optimiser which makes the final decision as to which Consensus Protocol to be selected and this decision is communicated back to the system resulting to a dynamic switch of the Consensus Protocol used by the Blockchain nodes.  

\section{EVALUATION}
\label{EVALUATION}

To demonstrate the suitability of the proposed approach, this section presents a quantitative analysis focusing on the optimisation of transaction latency by dynamically switching the Consensus Protocol. The analysis is based on a prototype implementation of the conceptual model presented in section \ref{AN INSTANTIATION FOR DYNAMIC CONSENSUS MANAGEMENT}.  The results obtained show that dynamically switching Consensus Protocols to reflect changes in the Blockchain system leads to better performance.

\subsection{Consensus Algorithms}
Two Consensus Protocols have been used for the experiments, the Istanbul Byzantine Fault Tolerance (IBFT)~\cite{moniz2020istanbul} and BigFoot~\cite{saltini2022bigfoot}, with IBFT having the ability to tolerate less stable network conditions and node failures and BigFoot being very efficient under stable ones. Both algorithms require 3f+1 nodes for tolerating f faulty nodes and can achieve consensus with 2f+1 replies which is the theoretical optimal~\cite{1/3tol}.

\begin{figure*}[!t]
\includegraphics[width=1\textwidth]{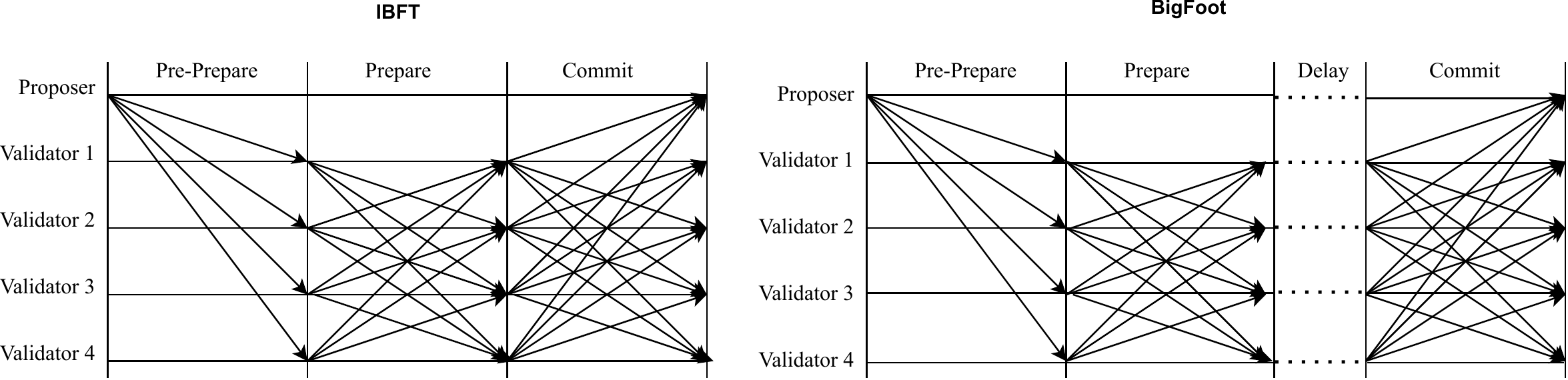}
\centering
\caption{The two Consensus Protocols used in the Analysis}
\label{Consensus}
\end{figure*}

The two protocols are illustrated in Figure \ref{Consensus} (in our analysis, block producer and validator are used interchangeably). BigFoot has two phases, namely \textit{Fast-path }and \textit{Fallback-path}. In the Fast-path, BigFoot is efficient requiring 2 message delays (namely Pre-prepare and Prepare) to reach consensus but 3f replies i.e. every node in the system must be online, in sync, and able to reply in time. To guarantee that BigFoot will eventually reach consensus even under less stable conditions, the Fast-path phase ends after a time out period if less than 3f but more that 2f replies are received (since the proposing node also counts as a validator,  2f replies imply 2f+1 validator votes). In this case, an extra Fallback-path phase is initiated which achieves consensus in 1 extra message delay (commit) and 2f+1 replies. PBFT on the other hand, always requires 3 message delays (pre-prepare, prepare, commit) and 2f+1 replies to reach consensus. 

It is evident from the above, that both algorithms sacrifice performance under certain conditions to excel in others, a fact, that the proposed Digital Twin approach can take advantage and dynamically switch betwee the two.  

\subsection{Simulation}
The conduct the experiments,  the BlockSim~\cite{blocksim} simulator was used and extended to support the modelling of  permissioned Blockchains and satisfy the requirements for the system. To the best of our knowledge, our extension of BlockSim is  the only one supporting dynamically changing the Consensus Protocols (IBFT and BigFoot) during runtime and one of only two tools supporting permissioned Blockchain simulation, the other being Talaria \cite{xing2021talaria}. Specifically, the Node, block, and Consensus modules of the BlockSim simulator were re-implemented to model a permissioned system. The Events structure of the system was changed from supporting high level events such as consensus and propagate block, to being able to model the lowest possible level events representing individual messages between the nodes to allow for more accurate modeling of the system. Finally, the Network module was augmented to model a unique network state for each node and allow for more complex network state to be modeled. 

Using BlockSim, a prototype model of the system illustrated  in figure \ref{arch} has been developed. The  block producers produce blocks of size $BS$ with a block interval of $BI$. 
The Consensus Protocol works in rounds. In each round a block producer is selected to propose a block, and initiate the consensus process by broadcasting the proposed block. After a block is accepted by a node, that node automatically advances to the next round. Each round has a timeout period after which the nodes initiate the round change process to agree on which round to advance to next. This guarantees the system's liveliness under faulty or malicious block producers. Finally, the Blockchain node syncing protocol is modified to include the current Consensus Protocol along with the new blocks. 

Data are provided to the Digital Twin periodically in time intervals ($TI$) of equal length. The system remains stable for a number of $TI$, and state changes occur every $TS$ ($TS$ being an integral multiple of $TI$). The parameters used in the system are the following: (a) no. of block producers, (b) state of block producers over time, (c) network state, (d) state of network over time, (e) avg. transactions per second, (f) transaction size, (g) max block size, (h) min block interval, and (i) round timeout.

In the simulation phase,  the scenario simulates the two  protocols for the next  time interval $TI_{n+1}$ and the results of the simulation i.e. the blocks produced, are fed to the optimiser. The optimiser selects the Consensus Protocol that yields the best average transaction latency. 

\subsection{Results}

The physical system model was designed to represent the worst case scenario, with faulty nodes periodically going offline, an unstable network, and a large number of transactions overloading the system. The model consists of ten block producers with two of them  being faulty. The system fluctuates between states every 100 time steps ($TS=100$), with the updates occurring every 25 time steps ($TI=25$). 

For performance evaluation, three metrics were used: (a) average transaction latency (b) average inter block time and (c) throughput. The average transaction latency was measured for each new block B added as follows 

\begin{equation}
    \frac{\sum_{i}^{T_B}{Time_B - Time_{T_i}}}{T_B}
\end{equation}

with $T_B$ denoting the number of transactions in B, $T_B$ the number of transactions included in B, $T_i$ the $i_{th}$ transaction in B and $Time_B$, $Time_{T_i}$ the time the block B and transaction $T_i$ were added to the system, respectively. The inter block time was measured as follows

\begin{equation}
    \frac{\sum_{i}^{B_{BC}}{Time_{B_i} - Time_{B_{i-1}}}}{B_{BC}}
\end{equation}
with $B_{BC}$ denoting the number of blocks. Finally, the throughput is defined as the number of transactions that the system is able to process per second and is defined as

\begin{equation}
    \frac{\sum_{i}^{B_{BC}}{T_{B_i}}}{T}
\end{equation}
with $T$ denoting the total system runtime.

The Digital Twin was tasked with optimising the above system by dynamically switching between the two consensus protocols (IBFT and BigFoot) and for comparison two other identical simulations were executed, one using the IBFT protocol and the second the BigFoot without any protocol changes. Specifically, the parameter values for the evaluation are the following: (a) no. of block producers is set to 10 (BP=10), (b) 2 faulty nodes periodically going offline (f=2), (c) network state ranges from 0.7MB/s to 2MB/s over time (e) avg. transactions per second is set to 50T/s, (f) transaction size is set to 5KB (g) max block size is set to 1MB (h) min block interval is set to 0.1s, and (i) the round timeout is set to 10s.

\begin{figure}[t]
    \centering
    \begin{subfigure}[b]{0.32\textwidth}
         \centering
         \includegraphics[width=1.2\textwidth]{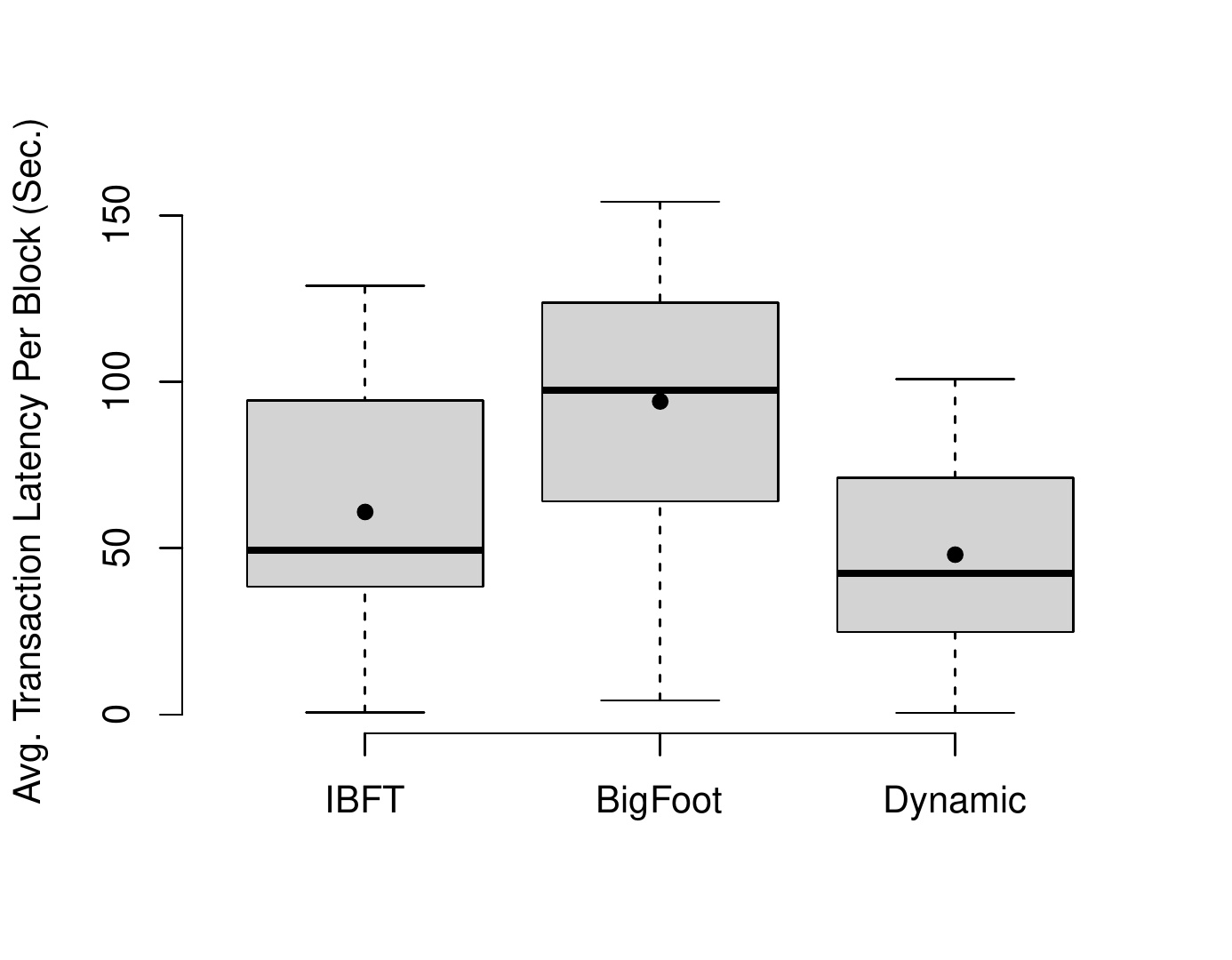}
         \caption{Latency (BP=10, f=2)}
         \label{fig:latency}
     \end{subfigure}
     \hfill
     \begin{subfigure}[b]{0.32\textwidth}
         \centering
         \includegraphics[width=1.2\textwidth]{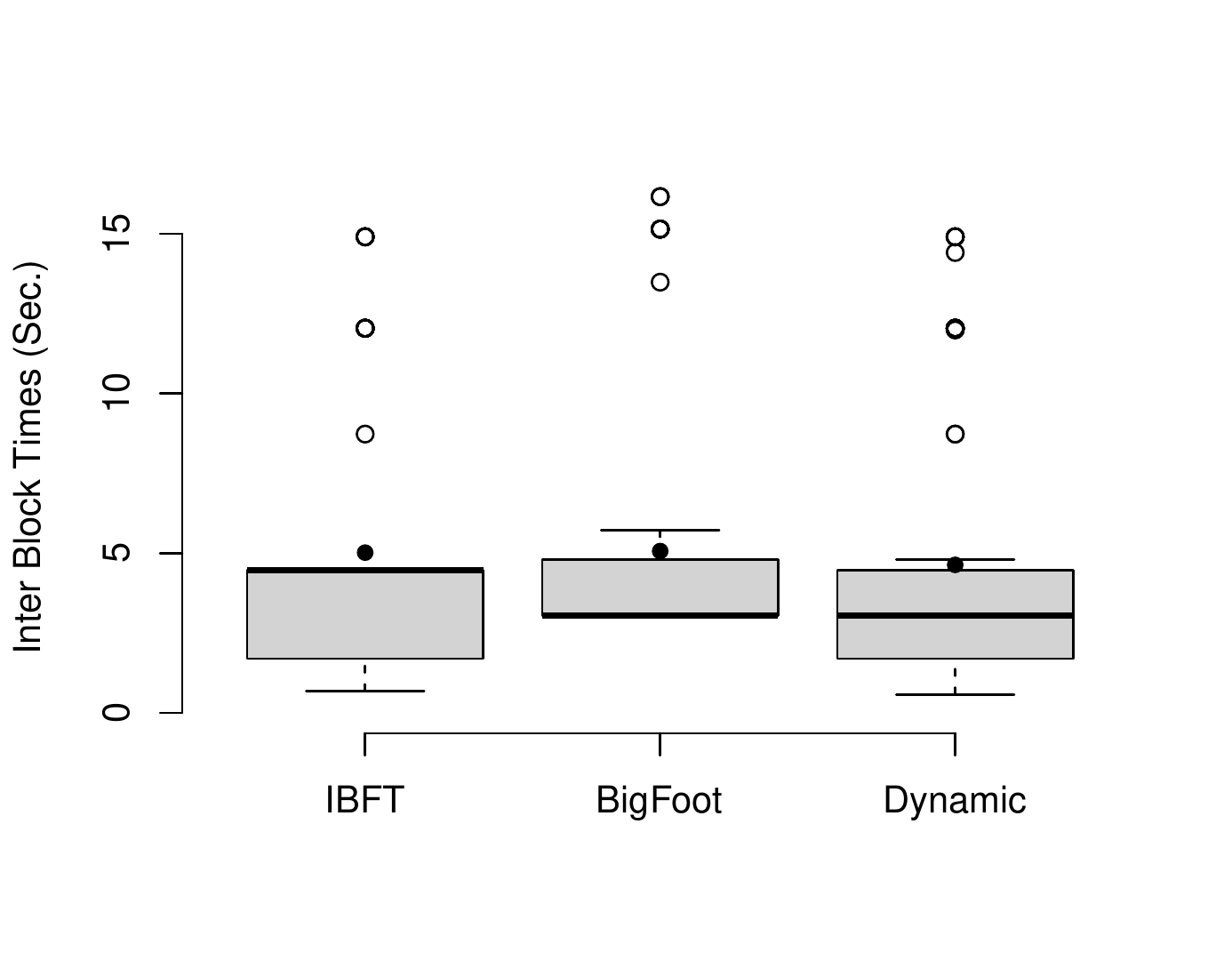}
         \caption{Inter Block Times (BP=10, f=2)}
         \label{fig:block times}
     \end{subfigure}
     \hfill
     \begin{subfigure}[b]{0.32\textwidth}
         \centering
         \includegraphics[width=1.2\textwidth]{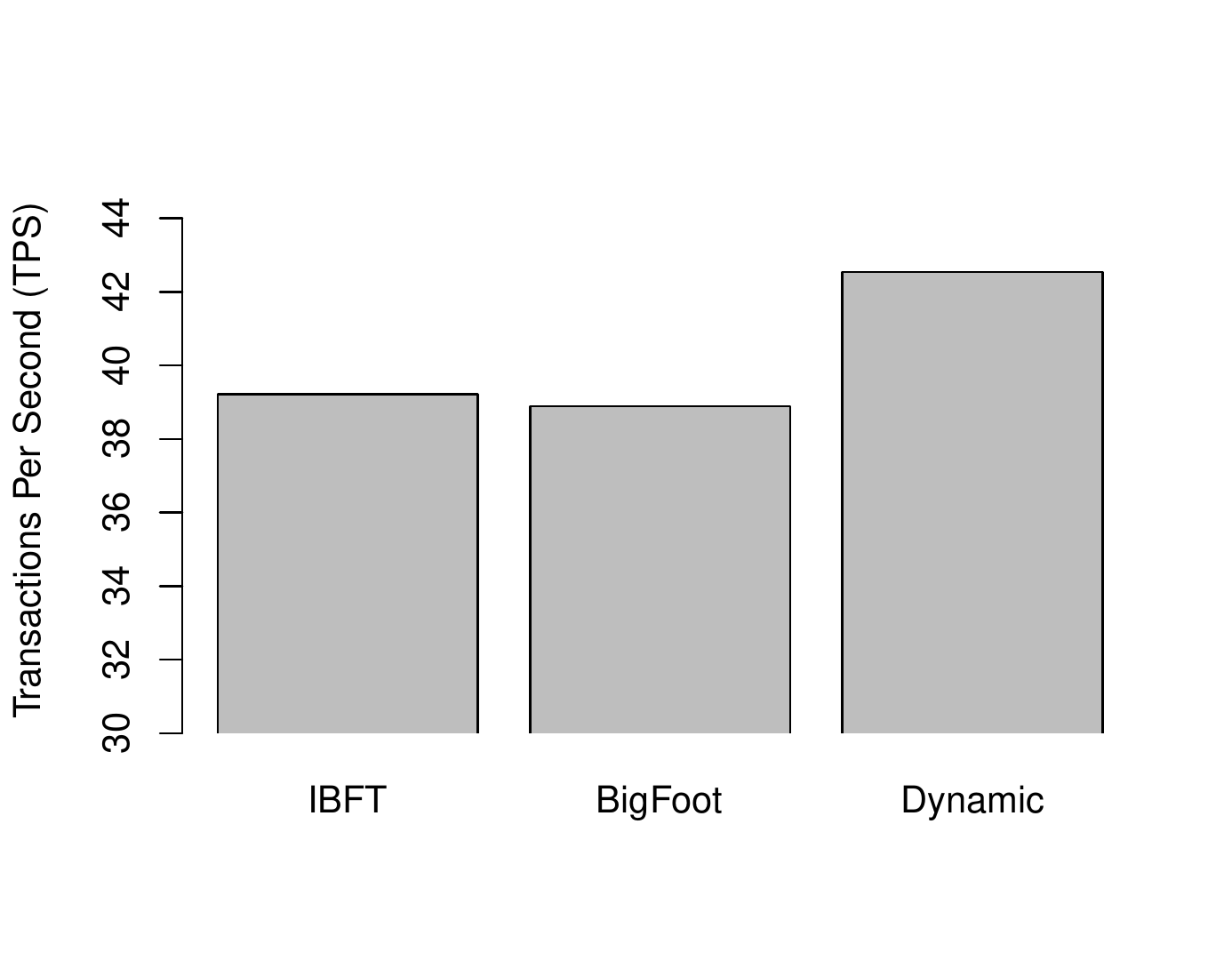}
         \caption{Throughput (BP=10, f=2)}
         \label{fig:troughput}
     \end{subfigure}
 
        \caption{Performance Results}
        \label{Results}
\end{figure}

Figure  \ref{Results} depicts the results of the experiments (note that black points in \ref{fig:latency} and \ref{fig:block times} denote the mean values), with Dynamic denoting the system optimised by the Digital Twin. It is evident that the dynamic protocol switching delivers the best performance. It achieves lower average transaction latency and inter block times, as well as higher throughput. These results confirm that a dynamic management of the blockchain by means of a Digital Twin is a viable approach to optimising a blockchain system by adapting the system parameters to reflect system and workload changes.

\section{CONCLUSIONS}
\label{CONCLUSIONS}

This paper has put forward the idea of utilising Digital Twins for the dynamic management of Blockchain systems in addressing the trilemma tradeoff. It has proposed a generic reference architecture and has demonstrated how the architecture can be instantiated 
for the dynamic selection and management of consensus in Blockchain-based systems to optimise performance, as core influencer to this tradeoffs. The experimental analysis has indicated that a Digital Twin can serve as a viable approach for dynamically managing the Trilemma tradeoff and help in improving performance.

Future work will further develop the architecture and refine it to dynamic analysis and management of richer set of scenarios and complex time varying tradeoffs. We will further extend the architecture to incorporate multiple views, each abstracting finer aspects of the trilemma. A more sophisticated optimiser will be developed that will utilise dynamic many optimisation and machine learning techniques to optimise within and across views to assist in planning and what-if analysis. For the what-if analysis, we will utilise Distributed Simulation techniques to scale the analysis at digital world, and achieve richer and faster than real time simulation.  
The info-symbiotic feedback loop will also be automated with bidirectional updates.

\bibliographystyle{wsc.bst}
\bibliography{ref.bib}

\end{document}